# Electric-field control of magnetic ordering in the tetragonal BiFeO$_3$

Hang-Chen Ding,[1] Chun-Gang Duan[1,2,*]

[1]*Key Laboratory of Polar Materials and Devices, Ministry of Education, East China Normal University, Shanghai 200062, China*

[2]*National Laboratory for Infrared Physics, Chinese Academy of Sciences, Shanghai 200083, China*

We propose a way to use electric-field to control the magnetic ordering of the tetragonal BiFeO$_3$. Based on systematic first-principles studies of the epitaxial strain effect on the ferroelectric and magnetic properties of the tetragonal BiFeO$_3$, we find that there exists a transition from C-type to G-type antiferromagnetic (AFM) phase at in-plane constant $a \sim 3.905$ Å when the ferroelectric polarization is along [001] direction. Such magnetic phase transition can be explained by the competition between the Heisenberg exchange constant $J_{1c}$ and $J_{2c}$ under the influence of biaxial strain. Interestingly, when the in-plane lattice constant enlarges, the preferred ferroelectric polarization tends to be canted and eventually lies in the plane (along [110] direction). It is found that the orientation change of ferroelectric polarization, which can be realized by applying external electric-field, has significant impact on the Heisenberg exchange parameters and therefore the magnetic orderings of tetragonal BiFeO$_3$. For example, at $a \sim 3.79$ Å, an electric field along [111] direction with magnitude of 2 MV/cm could change the magnetic ordering from C-AFM to G-AFM. As the magnetic ordering affects many physical properties of the magnetic material, e.g. magnetoresistance, we expect such strategy would provide a new avenue to the application of multiferroic materials.

PACS numbers: 75.85.+t, 75.10.-b, 77.55.Nv

The ability to possess (anti)ferromagnetism and ferroelectricity simultaneously renders multiferroic materials as promising candidates for future smart, portable and multifunctional microelectronic devices [1]. Most single phase multiferroic materials found up to now, e.g. bismuth ferrite (BiFeO$_3$, BFO), however, are antiferromagnets (or very weak ferromagnets) which are insensitive to the applied magnetic field, as the superexchange effect dominates the magnetic interactions in those oxides [2,3]. This puts a severe limitation on the application of single phase multiferroic materials. So far the successful application of BFO involving its multiferroic property is limited to exchange bias effect [4,5,6]. There are also reports using BFO as insulating barrier in spintronic devices, specifically magnetic tunnel junctions [7]. Nevertheless, at this stage such junctions have not utilized the ferroelectric property of BFO. Apparently, exploring new mechanism to the electric field control of the magnetic property would be of great interest to the application of BFO and other antiferromagnetic (AFM) multiferroic materials.

In this Letter, we propose an alternative way to utilize the multiferroic property of BFO. We demonstrate that the magnetic orderings of tetragonal BFO films can be altered by applying external electric field. Therefore many important physical properties related to magnetic ordering can be affected by electric field. For examples, the magnetoresistance is sensitive to the magnetic orderings of the material. As a consequence such strategy then provides a convenient way to "write" and "read" bit information stored in BFO based spintronic devices. The idea of using magnetic ordering to affect the physical properties of magnetic materials is indeed not novel. It is, however, hard to be realized in practical applications, as the magnetic ordering transition generally occurs under the change of temperature, strain (pressure), and impurity doping, *etc.*, which are all not appropriate to fast and repeatable control of magnetic orderings. Apparently, switching between different magnetic orderings using electric field has great advantage over other methods, and is actually another type of electric field control of magnetism [5,8-12]. As we will show in detail later, the electric field strategy is made possible only through the fascinating magnetoelectric couplings in BFO, which demonstrates a rich phase diagram and can be manipulated by many parameters.

We focus our study on the BFO with tetragonal structure. Previous first-principles calculations using the local spin-density approximation predict that tetragonal phase of BFO is less energetic favorable than the rhombohedral phase [13]. Modern thin film fabrication techniques, however, offer powerful ways to assemble metastable materials through epitaxial stabilization with a wide range of flexibility [14], including BFO [15]. We are then encouraged to study the effect of epitaxial strain, from compressive to tensile, on the physical properties of tetragonal BFO.

Our strategy is first to establish the relationship between the magnetic ground state and the epitaxial strain in a fixed ferroelectric polarization direction. Then we change the polarization directions, e.g. from perpendicular to parallel to the film plane, and repeat the above procedure. If at the same in-plane lattice constant, the magnetic orderings of the ground states are different for different polarizations, then it is *possible* to change the magnetic orderings of the system by altering the ferroelectric polarizations. Note here though traditionally the electric field on the ferroelectric film is applied perpendicularly to the film plane, present technology can make it possible to apply in-plane electric field to horizontally switch the polarization, as demonstrated in Ref. [5], and making the polarization to be canted is even easier [16].

The formation of different magnetic orderings is generally due to the delicate balance between the exchange interactions of the neighboring magnetic ions. Therefore we carry out a systematic study on the exchange interactions of tetragonal BFO using the Heisenberg Hamiltonian:

$$H = -\sum_n J_n \sum_{i>j} S_i \cdot S_j, \quad (1)$$

where $S_i$ is the spin-vector on site $i$, $J_n$ is the exchange constant and $n$ is the index of the nearest-neighbor shell. In this study, we count the exchange effect up to the second nearest-neighbor. In tetragonal structure, as in-plane lattice constant $a$ generally is different from out-of-plane lattice constant $c$, therefore there are four different exchange constants, i.e., $J_{1a}$, $J_{1c}$, $J_{2a}$ and $J_{2c}$, as shown in Fig. 1.

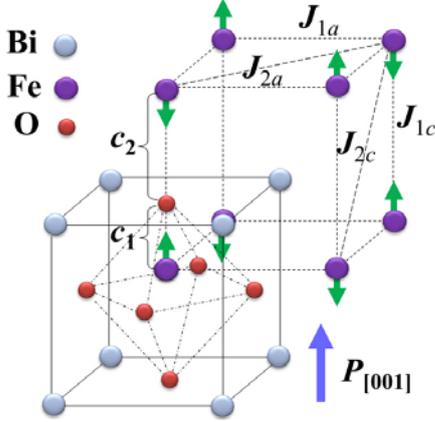

FIG. 1 (color online). Schematic of tetragonal BiFeO$_3$ (BFO) with G-type antiferromagnetic ordering and [001] polariztion. The spins of neighboring Fe atoms are shown as up and down arrows. Four different exchange constants: $J_{1a}$, $J_{1c}$, $J_{2a}$ and $J_{2c}$ are labeled accordingly. $c_1$ and $c_2$ denote the distances between the intermediate O and two neighboring Fe ions forming superexchange interaction along the $c$ direction.

Basically, there are four kinds of AFM orderings for the cubic structure, i.e. A, C, G and E-type AFM [17]. The AFM ordering can generally be described as alternative ferromagnetic (FM) sheets antiferromagnetically coupled along the sheet normal direction. For example, the ferromagnetic sheets for cubic structures are (001), (110) and (111) planes for A-, C- and G-AFM, respectively. The E-AFM can be regarded as a mixture of A- and C-AFM. For tetragonal structure, again due to the difference between $a$ and $c$, there are two kinds of magnetic orderings for A, C and E-type AFM, respectively. As a result, we consider totally eight magnetic orderings in the present study, including ferromagnetic and seven AFM orderings, namely A1, A2, C1, C2, E1, E2 and G. Following previous strategy [18], we can then compute the total energies of the above magnetic states using first-principles method, and extract the exchange parameters from solving a series of Eq. (1). Note that to avoid any calculation error caused by different symmetries or sizes of the unit cell, all calculations are performed using a $2 \times 2 \times 2$ supercell which contains eight BFO formula units (40 atoms).

For all the calculations presented in this work we use the highly accurate full potential projector augmented wave method as implemented in the Vienna *ab initio* Simulation Package (VASP) [19]. We employ the Perdew-Burke-Ernzerhof (PBE) form [20] of the generalized gradient approximation (GGA) for the exchange and correlation. The effective Hubbard constant $U_{\text{eff}} = U - J = 2$ eV [21] is adopted to treat the strongly-correlated nature of BFO. We use the energy cutoff of 500 eV for the plane wave expansion of the PAWs and a $4 \times 4 \times 4$ Monkhorst-Pack grid for k-point sampling in the total energy calculations for the suprcell. At each lattice constant and polarization direction, the structure of BFO, including the atomic coordinates and $c/a$ ratio, are optimized until the Hellman-Feynman force on each atom becomes less than 2 meV/Å. The Berry phase technique is used to calculate ferroelectric polarizations [22].

We first consider the case of [001] polarization. Under this polarization direction, the relaxed cell is strongly distorted from the cubic perovskite structure and forms a so called *super tetragonal* structure with a giant $c/a$ ratio [23,24]. Our calculations indicate that the strain effect has significant influence on the $c/a$ ratio. For instance, when the lattice constant increases from 3.65 to 4.2 Å, the $c/a$ ratio drops rapidly from 1.37 to 0.93. Interestingly, the ferroelectric polarization does not change dramatically with the strain effect, and keeps a value around 1.5 C/m$^2$. These results are in good agreement with previous studies [23-25]. In the whole range we studied, it is shown that the competition for the ground state is between C1 and G-type AFM, and other magnetic states are at least 40 meV per formula unit higher in energy than the ground state. This is due to the strong superexchange interactions between the nearest-neighbors in the *ab* plane, therefore A and E-type AFM, where nearest-neighbors are ferromagnetic coupled, are excluded from the competition. The magnetic ground state is at $a \sim 3.74$ Å with C1-AFM ordering, and the C1-G magnetic transition occurs at around 3.905 Å, as shown in Fig. 2.

To find out the driving mechanism of this C1-G transition, we write down the energy difference between the C1 and G states using Eq. (1) and obtain

$$\begin{aligned}\Delta J^{[001]} &= H_{C1} - H_G \\ &= (2J_{1a} - J_{1c} - 2J_{2a} + 4J_{2c}) - (2J_{1a} + J_{1c} - 2J_{2a} - 4J_{2c}) \\ &= -2J_{1c} + 8J_{2c} \end{aligned} \quad (2)$$

As is clear from Eq. (2), it is the competition between $J_{1c}$, which generally has larger magnitude, and $J_{2c}$, which has more numbers, that determines the magnetic ground state. In Fig. 2, we plot the change of $J_{1c}$, $J_{2c}$ and $\Delta J^{[001]}$ with the in-plane lattice constant. Note that we have more equations than variables to be solved, and this provides a way to check the consistency of our calculated exchange parameters. Indeed we find that our calculated values for $J_{1a}$, $J_{1c}$, $J_{2a}$ and $J_{2c}$ change only very slightly on the choices of the equations to be solved. On the contrary, if we only consider the exchange interactions in the nearest-neighbor shell ($J_{1a}$ and $J_{1c}$), the results are not consistent at all. This suggests that it is important to include the contribution from the second nearest-neighbors in the evaluation of magnetic interactions. From Fig. 2, we can see that both $J_{1c}$ and $J_{2c}$ are negative and are of almost the same magnitude for small $a$. As $J_{2c}$ wins by numbers, the system is in C1-AFM state. Nevertheless, the magnitude of $J_{1c}$ increases dramatically when the in-plane lattice constant $a$ increases, whereas $J_{2c}$ changes little with $a$. As a consequence, there occurs a C1-G AFM transition at around 3.905 Å.

The abrupt change with in-plane lattice constant of $J_{1c}$, i.e. the magnetic exchange strength between the two neighboring Fe ions along $c$ direction, can be understood by the significant change of Fe-O distances along the $c$ direction, especially $c_2$ which describes the bond length of the weakly coupled Fe-O ions, as is clearly from Fig. 2. For small in-plane lattice constant $a$, the tetragonality ($c/a$) is large and $c_2$ is much larger than regular Fe-O bond length. This results in significantly weak superexchange interaction between the neighboring Fe atoms along $c$ direction, as the Fe-O hopping parameter $t_{pd}$ strongly depends on the Fe-O bond lengths. When $a$ increases, both the $c/a$ ratio and $c_2$ drop quickly and the superexchange effect along $c$ direction becomes much stronger. As a comparison, $J_{2c}$ changes little with $a$, possibly because the distance between two Fe atoms on the face diagonal [101] or [011] changes much slower than $c$. Then, when $a > 3.905$ Å, $\Delta J^{[001]}$ becomes positive, indicating the system transforms from C1 to G phase.

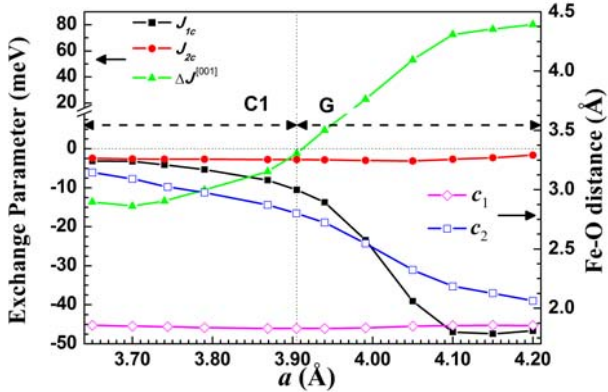

FIG. 2 (color online). Evolution of the Heisenberg exchange parameters $J_{1c}$ (solid square), $J_{2c}$ (solid circle), and $\Delta J^{[001]}$ (triangle). Curves $c_1$ (open diamond) and $c_2$ (open square) are the distances of Fe-O atoms which form the superexchange interaction along the $c$ direction, see Fig. 1 for illustrated definition.

In the above studies, we assume the polarization is along (001) direction when the in-plane lattice constant increases. This is, however, no longer true for the ground state of tetragonal BFO under biaxial tensile strain, as the in-plane or canted ferroelectric instability will be allowed to develop when the in-plane lattice constant increases. Therefore we repeat similar studies on the tetragonal BFO on polarizations along [111] and [110] directions. Note here we actually have calculated several directions within the $ab$ plane and find that the polarization along [110] (and its equivalent) directions is the most stable state in the plane, agreeing with experiments [26]. The change of polarization direction indeed will reduce the original tetragonal symmetry $P4mm$. Nevertheless we assume the in-plane clamping is strong enough to keep the tetragonal structure of the system, yet the $c/a$ ratio is allowed to change when ferroelectric polarization is canted or in-plane.

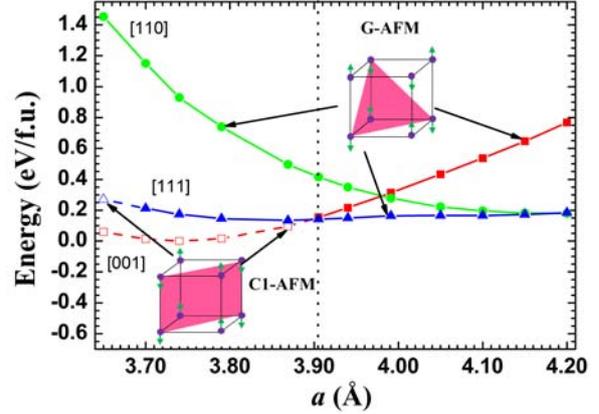

FIG. 3 (color online). Total energies, respect to the energy of [001] polarization at 3.74 Å, of the tetragonal BFO with ferroelectric polarization along [001] (square), [111] (triangle), and [110] (circle) directions. The straight lines with solid symbols and dotted lines with open symbols indicate the magnetic orderings are G- and C1-AFM, whose spin arrangements are shown as insets, respectively.

In Fig. 3, we can see that with the increase of in-plane lattice constant, state with [111] polarization gradually becomes energetically favorable and eventually defeats [001] polarization at around 3.89 Å. State with [110] polarization decreases its energy dramatically when $a$ increases from 3.65 to 3.905 Å, and becomes much more stable after that. When the biaxial tensile strain is large enough, it becomes the most stable state for all the polarization directions.

The interesting part of Fig. 3 is, for [111] polarized states the C1-G transition occurs at much smaller in-plane lattice constant ($a$~3.7Å). Detailed analysis reveals that polarization canting decreases $c/a$ ratio, and $J_{2c}$ is overwhelmed by $J_{1c}$ due to the reinforcement of the superexchange effect between the two neighboring Fe atoms along the $c$ direction, making G-type AFM energy favorable at smaller lattice constant. Therefore, in a quite large range (3.7~3.905 Å), it is possible to switch the magnetic ordering from C1-type to G-type AFM by changing the polarization from [001] to [111] direction. We estimate the magnitude of the required external electric-field at $a$ = 3.79 Å to be around 2 MV/cm, which is easy to be realized experimentally for ultrathin films. At larger $a$, the required electric field is even smaller.

The changes on the magnetic and electronic structures of tetragonal BFO due to the polarization switching can be more clearly seen in Fig. 4, which plots the partial charge densities from 1 eV below the Fermi level ($E_F$) to $E_F$ of the majority-spin electrons in the (010) plane of tetragonal BFO at 3.79 Å for [001] (C1-AFM, Fig. 4a) and [111] (G-AFM, Fig. 4b) polarizations. Note since both states are in AFM ordering, the minority-spin charge density plots can be obtained by shifting the corresponding majority-spin plot by in-plane lattice constant $a$ along $x$-axis. Along $z$ direction, the spin-dependent charge densities are significantly different for [001] and [111] polarized states. In C1-AFM, where the spins along $z$ direction are in FM order, the charge density changes continuously in one spin channel (either spin-up or spin-down), whereas for G-AFM, where the spins along $z$ direction are in AFM order, the charge density are discontinuous in both spin channels. This will inevitably affect the spin-transport properties of the BFO film.

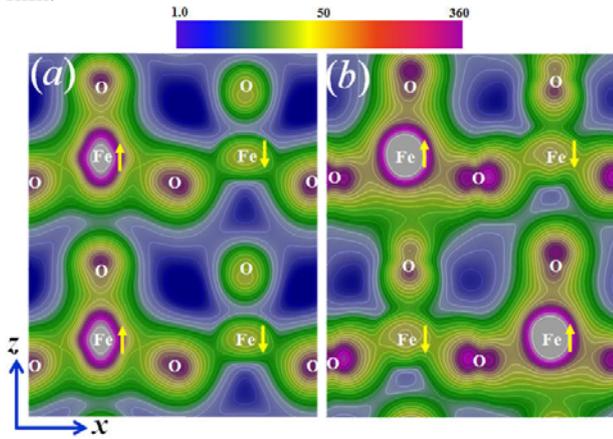

FIG. 4 (color online). Majority-spin charge density (in arbitrary units), calculated in the energy window from $E_F$-1 eV to $E_F$ in the (010) plane, of tetragonal BFO at 3.79 Å when ferroelectric polarization is along (a) [001] (C1-AFM) and (b) [111] direction (G-AFM). The up and down arrows on Fe atoms indicate the spin orientations.

For the in-plane polarization [110], the importance of $J_{2c}$ can be neglected when comparing with $J_{1c}$, due to the even smaller $c/a$ ratio when the polarization is in the plane. Therefore in a wide range of the in-plane lattice constant, $J_{1a}$ is dominant and remains to be negative. As a matter of fact, for [110] polarization G-AFM is always preferred from 3.65 to 4.20 Å (Fig. 3). Its total energy, however, is much higher than that of [111] polarized state when $a$ is smaller than 3.905 Å. Therefore the [110] polarized state requires much stronger, but still experimentally achievable, electric field to be switched from [001] polarized state.

In summary, our theoretical study demonstrates that electric field could be used to control the magnetic ordering of BFO with tetragonal structure. At certain range of in-plane lattice constant, i.e. 3.70~3.905 Å as predicted by our GGA calculation, applying canted external electric field could change the ferroelectric polarization direction, which would bring at least two effects to the magnetic interactions of the system. One is to change the $c/a$ ratio of the unit cell, the other is to change the bondings of Fe ions with neighboring O atoms. Both effects could result in significant change on the Heisenberg exchange parameters, which determine the magnetic orderings. The estimated electric field to change the magnetic orderings of tetragonal BFO is experimentally achievable, depending on the in-plane strain. We hope that our theoretical prediction of electric field control of magnetic ordering of BFO will stimulate further experimental studies.

We acknowledge fruitful discussions with Y.-H. Chu and Q. He. This work was supported by the NSF of China (Grant No. 50771072 and 50832003), PCSIRT, NCET, and 973 Program No. 2007CB924900. Computations were performed at the ECNU computing center.

*E-mail: wxbdcg@gmail.com or cgduan@clpm.ecnu.edu.cn.

4